\documentclass[12pt,english]{article}
\usepackage[T1]{fontenc}
\usepackage[latin1]{inputenc}
\usepackage{a4wide}
\usepackage{amssymb}

\makeatletter

\makeatletter

\@addtoreset{equation}{section}

\@addtoreset{figure}{section}

\@addtoreset{table}{section}
\makeatother

\usepackage{babel}
\makeatother
\begin{document}
\date{March 26, 2007}

\title{\begin{flushright}{\normalsize ITP-Budapest Report No. 631}\end{flushright}\vspace{1cm}Spectrum
of local boundary operators from boundary form factor bootstrap}

\author{M. Sz{\H o}ts$^{1}$ and G. Takács$^{2}$%
\thanks{E-mail: takacs@elte.hu%
}\\
$^{1}$Eötvös University, Budapest, Hungary\\
$^{2}$Theoretical Physics Research Group of the Hungarian Academy
of Sciences\\
}

\maketitle
\begin{abstract}
Using the recently introduced boundary form factor bootstrap equations,
we map the complete space of their solutions for the boundary version
of the scaling Lee-Yang model and sinh-Gordon theory. We show that
the complete space of solutions, graded by the ultraviolet behaviour
of the form factors can be brought into correspondence with the spectrum
of local boundary operators expected from boundary conformal field
theory, which is a major evidence for the correctness of the boundary
form factor bootstrap framework.
\end{abstract}

\section{Introduction}

The bootstrap program aims to classify and explicitly solve 1+1 dimensional
integrable quantum field theories by constructing all of their Wightman
functions. For bulk theories, the first stage is the S-matrix bootstrap:
the scattering matrix, connecting asymptotic \emph{in} and \emph{out}
states, is determined from its properties such as factorizability,
unitarity, crossing symmetry and the Yang-Baxter equation supplemented
by the maximal analyticity assumption which results in the complete
on-shell solution of the theory, i.e. the spectrum of excitations
and their scattering amplitudes (for reviews see \cite{Sbstr,Sanal}).
The second step is the form factor bootstrap, which allows one to
determine matrix elements of local operators between asymptotic states
(form factors) using their analytic properties originating from the
already known S-matrix. The form factors are then used to build the
correlation (Wightman) functions via their spectral representations,
yielding a complete off-shell description of the theory (see \cite{Smirnov}
for a review). 

The first step of an analogous bootstrap program for 1+1 dimensional
integrable \emph{boundary} quantum field theories, the boundary R-matrix
bootstrap, was developed in the pioneering work by Ghoshal and Zamolodchikov
\cite{GZ}; it makes possible the determination of the reflection
matrices and provides complete description of the theory on the mass
shell.

For the second step matrix elements of local operators between asymptotic
states have to be computed. In a boundary quantum field theory there
are two types of operators, the bulk and the boundary operators, where
their names indicate their localization point. The boundary bootstrap
program, namely the boundary form factor program for calculating the
matrix elements of local boundary operators between asymptotic states
was initiated in \cite{bffprogram}%
\footnote{There exists no analogous framework for bulk operators in the presence
of the boundary yet. Large distance expansion for their correlators
in terms of bulk form factors can be given using the boundary state
formalism, but so far this approach has only been used to compute
one-point functions \cite{Muss1pt,DPTW1pt}.%
}.

In this work we make further progress in understanding the boundary
form factor bootstrap by classifying and counting the solutions. In
the bulk case it was proposed by Cardy and Mussardo that the space
of solutions of form factor axioms is to be identified with the space
of local operators \cite{cardy_mussardo}. The program of counting
the solutions was performed for several interesting models \cite{koubek_mussardo,koubek1,koubek2,smirnovcounting}
and it was shown that the results agree with the spectrum of local
operators expected from the Lagrangian description in terms of the
ultraviolet limiting conformal field theory. Here we take the models
treated in \cite{bffprogram} and perform a similar counting of solutions,
which can then be brought into correspondence with the spectrum of
boundary operators predicted by boundary conformal field theory.

The outline of the paper is as follows. In Section 2 we give a short
overview of the boundary form factor axioms and also introduce some
basic notions concerning form factor solutions. In Section 3 we investigate
the scaling Lee-Yang model with boundary, which has two physically
inequivalent boundary conditions. We start with the simpler boundary
condition (denoted by $\mathbf{1}$) and develop all the concepts
necessary to perform the counting of the form factor solutions. Then
we apply these to the other boundary condition (denoted by $\Phi$)
too, while in Section 4 we treat the boundary sinh-Gordon model using
the same tools. We give our conclusions and discuss major open problems
in Section 5.

\section{The boundary form factor bootstrap and the classification of solutions}

\subsection{The boundary form factor axioms}

The axioms satisfied by the form factors of a local boundary operator
were derived in \cite{bffprogram}. Here we only list them without
much further explanation. Let us suppose that we treat an integrable
boundary quantum field theory in the domain $x<0$, with a single
scalar particle of mass $m$, which has a two-particle $S$ matrix
$S(\theta)$ (using the standard rapidity parametrization) and a one-particle
reflection factor $R(\theta)$ off the boundary, satisfying the boundary
reflection factor bootstrap conditions of Ghoshal and Zamolodchikov
\cite{GZ}. For a local operator $\mathcal{O}(t)$ localized at the
boundary (located at $x=0$, and parameterized by the time coordinate
$t$) the form factors are defined as\begin{eqnarray*}
\,_{out}\langle\theta_{1}^{'},\theta_{2}^{'},\dots,\theta_{m}^{'}\vert\mathcal{O}(t)\vert\theta_{1},\theta_{2},\dots,\theta_{n}\rangle_{in} & =\\
 &  & \hspace{-2cm}F_{mn}^{\mathcal{O}}(\theta_{1}^{'},\theta_{2}^{'},\dots,\theta_{m}^{'};\theta_{1},\theta_{2},\dots,\theta_{n})e^{-imt(\sum\cosh\theta_{i}-\sum\cosh\theta_{j}^{'})}\end{eqnarray*}
for $\theta_{1}>\theta_{2}>\dots>\theta_{n}>0$ and $\theta_{1}^{'}<\theta_{2}^{'}<\dots<\theta_{m}^{'}<0$,
using the asymptotic $in/out$ state formalism introduced in \cite{BBT}.
They can be extended analytically to other values of rapidities. With
the help of the crossing relations derived in \cite{bffprogram} all
form factors can be expressed in terms of the elementary form factors\[
\,_{out}\langle0\vert\mathcal{O}(0)\vert\theta_{1},\theta_{2},\dots,\theta_{n}\rangle_{in}=F_{n}^{\mathcal{O}}(\theta_{1},\theta_{2},\dots,\theta_{n})\]
which can be shown to satisfy the following axioms: 

I. Permutation:

\begin{center}\[
F_{n}^{\mathcal{O}}(\theta_{1},\dots,\theta_{i},\theta_{i+1},\dots,\theta_{n})=S(\theta_{i}-\theta_{i+1})F_{n}^{\mathcal{O}}(\theta_{1},\dots,\theta_{i+1},\theta_{i},\dots,\theta_{n})\]
\end{center}

II. Reflection:\[
F_{n}^{\mathcal{O}}(\theta_{1},\dots,\theta_{n-1},\theta_{n})=R(\theta_{n})F_{n}^{\mathcal{O}}(\theta_{1},\dots,\theta_{n-1},-\theta_{n})\]

III. Crossing reflection: \[
F_{n}^{\mathcal{O}}(\theta_{1},\theta_{2},\dots,\theta_{n})=R(i\pi-\theta_{1})F_{n}^{\mathcal{O}}(2i\pi-\theta_{1},\theta_{2},\dots,\theta_{n})\]

IV. Kinematical singularity\[
-i\mathop{\textrm{Res}}_{\theta=\theta^{'}}F_{n+2}^{\mathcal{O}}(\theta+i\pi,\theta^{'},\theta_{1},\dots,\theta_{n})=\left(1-\prod_{i=1}^{n}S(\theta-\theta_{i})S(\theta+\theta_{i})\right)F_{n}^{\mathcal{O}}(\theta_{1},\dots,\theta_{n})\]
or equivalently described as \[
-i\mathop{\textrm{Res}}_{\theta=\theta^{'}}F_{n+2}^{\mathcal{O}}(-\theta+i\pi,\theta^{'},\theta_{1},\dots,\theta_{n})=\left(R(\theta)-\prod_{i=1}^{n}S(\theta-\theta_{i})R(\theta)S(\theta+\theta_{i})\right)F_{n}^{\mathcal{O}}(\theta_{1},\dots,\theta_{n})\]

V. Boundary kinematical singularity

\[
-i\mathop{\textrm{Res}}_{\theta=0}F_{n+1}^{\mathcal{O}}(\theta+\frac{i\pi}{2},\theta_{1},\dots,\theta_{n})=\frac{g}{2}\Bigl(1-\prod_{i=1}^{n}S\bigl(\frac{i\pi}{2}-\theta_{i}\bigr)\Bigr)F_{n}^{\mathcal{O}}(\theta_{1},\dots,\theta_{n})\]
where $g$ is the one-particle coupling to the boundary\begin{equation}
R(\theta)\sim\frac{ig^{2}}{2\theta-i\pi}\quad,\quad\theta\sim i\frac{\pi}{2}\label{eq:gdef}\end{equation}

VI. Bulk dynamical singularity \[
-i\mathop{\textrm{Res}}_{\theta=\theta^{'}}F_{n+2}^{\mathcal{O}}(\theta+iu,\theta^{'}-iu,\theta_{1},\dots,\theta_{n})=\Gamma F_{n+1}^{\mathcal{O}}(\theta,\theta_{1},\dots,\theta_{n})\]
corresponding to a bound state pole of the $S$ matrix\[
S(\theta)\sim\frac{i\Gamma^{2}}{\theta-2iu}\quad,\quad\theta\sim2iu\]
(in a theory with a single particle, the only possible value is $u=\pi/3$).

VII. Boundary dynamical singularity

\[
-i\mathop{\textrm{Res}}_{\theta=iv}F_{n+1}^{\mathcal{O}}(\theta_{1},\dots,\theta_{n},\theta)=\tilde{g}\tilde{F}^{\mathcal{O}}(\theta_{1},\dots,\theta_{n})\]
which corresponds to a pole in the reflection factor describing a
boundary excited state:\[
R(\theta)\sim\frac{i\tilde{g}^{2}/2}{\theta-iv}\quad,\quad\theta\sim iv\]
We further assume \emph{maximum analyticity} i.e. that the form factors
have only the minimal singularity structure consistent with the above
axioms.

\subsection{Solution of the axioms}

\subsubsection{The general Ansatz}

The general form factor solution can be written in the following form
\cite{bffprogram}\begin{equation}
F_{n}(\theta_{1},\theta_{2},\dots,\theta_{n})=G_{n}(\theta_{1},\theta_{2},\dots,\theta_{n})\prod_{i=1}^{n}r(\theta_{i})\prod_{i<j}f(\theta_{i}-\theta_{j})f(\theta_{i}+\theta_{j})\label{eq:GenAnsatz}\end{equation}
where $f$ is the \emph{minimal bulk two-particle form factor (2PFF)}
satisfying the conditions\[
f(\theta)=S(\theta)f(-\theta),\qquad f(i\pi+\theta)=f(i\pi-\theta)\]
and having the minimum possible number of singularities in the physical
strip $0\leq\theta<\pi$ together with the slowest possible growth
at infinity \cite{KW}, and $r$ is the \emph{minimal boundary one-particle
form factor (1PFF)} satisfying\begin{equation}
r(\theta)=R(\theta)r(-\theta)\quad;\quad r(i\pi+\theta)=R(-\theta)r(i\pi-\theta)\label{eq:1pff}\end{equation}
plus analytic conditions similar to those of $f$, but in this case
in the strip $0\leq\theta<\pi/2$. 

The functions $G_{n}$ are totally symmetric and meromorphic in the
rapidities $\theta_{i}$. They are also even and periodic in them
with the period $2\pi i$, so they can only be functions of the variables\[
y_{i}=\mathrm{e}^{\theta_{i}}+\mathrm{e}^{-\theta_{i}}\]
Let us now turn to the analysis of the singularity structure. In a
theory with only one particle, the only possible singularity of the
$S$ matrix in the physical strip is located at $\theta=2\pi i/3$
corresponding to the self-fusion of the particle (plus the crossed
channel pole for the same process at $\pi i/3$) and the relevant
fusion coupling is defined as\[
\Gamma^{2}=-i\mathop{\mathrm{Res}}_{\theta=\frac{2\pi}{3}i}S(\theta)\]
We suppose that the 2PFF function $f$ is chosen such that it has
a pole at $\theta=2\pi i/3$ so that it encodes this singularity (for
an example see (\ref{eq:min2pffly})). We further assume that the
boundary dynamical singularities (but not the kinematical ones!) are
similarly contained in the 1PFF function $r$. Then the functions
$G_{n}$ only have singularities at the positions of the kinematical
singularities, and so they can be written in the form\[
G_{n}(\theta_{1},\theta_{2},\dots,\theta_{n})=\frac{Q_{n}(y_{1},y_{2}\dots,y_{n})}{\prod_{i}y_{i}\,\prod\limits _{i<j}(y_{i}+y_{j})}\]
where the $Q_{n}$ are entire functions symmetric in their arguments.

\subsubsection{Two-point functions and scaling weights}

The two-point functions can be written using a spectral representation:\begin{equation}
\langle0\vert\mathcal{O}(t)\mathcal{O}(0)\vert0\rangle=\sum_{n=0}^{\infty}\frac{1}{(2\pi)^{n}}\int_{\theta_{1}>\theta_{2}>\dots>\theta_{n}>0}d\theta_{1}d\theta_{2}\dots d\theta_{n}\, F_{n}F_{n}^{+}\exp\left(-imt{\displaystyle \sum_{i=1}^{n}}\cosh\theta_{i}\right)\label{eq:2pt}\end{equation}
where time translation invariance was used and the form factors $F_{n}$,
$F_{n}^{+}$ are defined as\[
F_{n}=\langle0\vert\mathcal{O}(0)\vert\theta_{1},\theta_{2},\dots,\theta_{n}\rangle_{in}=F_{n}^{\mathcal{O}}(\theta_{1},\theta_{2},\dots,\theta_{n})\]
and\[
F_{n}^{+}=\,_{in}\langle\theta_{1},\theta_{2},\dots,\theta_{n}\vert\mathcal{O}(0)\vert0\rangle=F_{n}^{\mathcal{O}}(i\pi+\theta_{n},i\pi+\theta_{n-1},\dots,i\pi+\theta_{1})\]
which, for unitary theories, is the complex conjugate of the previous
one: $F_{n}^{+}=F_{n}^{*}$. In the Euclidean time $\tau=it$ the
form factor expansion of the correlator converges rapidly for large
separations since multi-particle terms are exponentially suppressed.

We are interested in operators which can be classified according to
their scaling dimensions, which means that the two-point function
must have a power-like short-distance singularity\begin{equation}
\langle0\vert\mathcal{O}(\tau)\mathcal{O}(0)\vert0\rangle=\frac{1}{\tau^{2\Delta}}\label{eq:shortdistance}\end{equation}
where $\Delta$ is an exponent determined by the ultraviolet scaling
weights of the local fields. More precisely let us consider the conformal
operator product expansion\[
\mathcal{O}(\tau)\mathcal{O}(0)\sim\sum_{h_{i}}\frac{C_{\mathcal{OO}}^{i}}{\tau^{2h-h_{i}}}\mathcal{O}_{i}(0)\]
where $h$ is the ultraviolet weight of the field $\mathcal{O}$ and
the $h_{i}$ are the weight of the $\mathcal{O}_{i}$. It is obvious
that\[
2\Delta=2h-h_{\mathrm{min}}\]
where $h_{\mathrm{min}}$ is the minimum of the weights of the operators
appearing in the expansion. In many cases it is $h_{\mathrm{min}}=0$
(corresponding to the weight of the identity) and then $h$ and $\Delta$
are identical.

It is well-known that the presence of a power-like singularity means
that the form factors themselves can only grow exponentially in the
rapidity variables \cite{delfino_mussardo} and so the functions $Q_{n}$
are symmetric polynomials in the variables $y_{i}$. The singularity
axioms give recursion relations for the polynomials $Q_{n}$, for
which we introduce the abbreviated notation\begin{eqnarray}
Q_{n} & = & \mathcal{K}\left[Q_{n+2}\right]\nonumber \\
Q_{n} & = & \mathcal{D}\left[Q_{n+1}\right]\nonumber \\
Q_{n} & = & \mathcal{B}\left[Q_{n+1}\right]\label{eq:genrecrel}\end{eqnarray}
where $\mathcal{K}$, $\mathcal{D}$ and $\mathcal{B}$ denote the
recursion relation resulting from the bulk kinematical, bulk dynamical
and boundary kinematical singularity axioms respectively. We give
explicit forms of these relations for specific models later.

If the form factors grow asymptotically as\begin{equation}
\left|F_{n}(\theta_{1}+\theta,\theta_{2}+\theta,\dots,\theta_{n}+\theta)\right|\sim\mathrm{e}^{x\left|\theta\right|}\quad\mathrm{as}\quad\left|\theta\right|\rightarrow\infty\label{eq:xdef}\end{equation}
(where we assume that the exponent $x$ is independent of the level
$n$)%
\footnote{For all models considered here we show later that the space of solutions
is spanned by so-called 'simple towers' for which the asymptotic exponent
$x$ is indeed independent of the level $n$.%
} then the leading short-distance behaviour of the individual terms
can be easily shown to be $1/\tau^{2x}$ and so naively $\Delta=x$.
In fact, leading logarithmic corrections in $\tau$ can (and in many
cases do) sum up to an anomalous contribution to the ultraviolet exponent
$\Delta$ and therefore we call $x$ the naive (or engineering) dimension
of the form factor solution.

\subsubsection{Towers and local operators}

Following the ideas in the work of Koubek and Mussardo on classification
of the \emph{bulk} form factor solutions \cite{koubek_mussardo},
the set of elementary form factors corresponding to a local operator
forms a \emph{tower} of form factors, graded by the number of particles:\[
\mathcal{O}\,\mapsto\,\mathcal{F}_{\mathcal{O}}=\left\{ F_{n}^{\mathcal{O}}(\theta_{1},\dots,\theta_{n})\right\} _{n\in\mathbb{N}}\]
Due to the recursion relations (\ref{eq:genrecrel}) the solutions
(\ref{eq:GenAnsatz})\[
F_{n}(\theta_{1},\theta_{2},\dots,\theta_{n})=\frac{Q_{n}(y_{1},y_{2}\dots,y_{n})}{\prod_{i}y_{i}\,\prod\limits _{i<j}(y_{i}+y_{j})}\prod_{i=1}^{n}r(\theta_{i})\prod_{i<j}f(\theta_{i}-\theta_{j})f(\theta_{i}+\theta_{j})\]
of the form factor equations can also be classified into towers which
consist of a single form factor at each level $n$ such that the solutions
at different levels are linked together via the relations (\ref{eq:genrecrel}).
As we show later in the explicit examples, the recursion relations
are such that the naive scaling dimension $x$ is independent of the
level $n$ and therefore it can be assigned to the tower itself. Therefore
the space of form factor solutions can be rearranged into a space
of towers $\mathcal{T}$, which can in turn be graded by the naive
scaling dimension. Due to the linearity of the form factor axioms,
the space of towers is a linear space i.e. whenever $\mathcal{F}=\left\{ F_{n}\right\} _{n\in\mathbb{N}}$and
$\mathcal{F}'=\left\{ F_{n}'\right\} _{n\in\mathbb{N}}$ are two towers,
so is their general linear combination\[
\alpha\mathcal{F}+\beta\mathcal{F}'=\left\{ \alpha F_{n}+\beta F_{n}'\right\} _{n\in\mathbb{N}}\]
While the space of towers ($\mathcal{T}$) is always infinite dimensional,
the subspaces corresponding to a given value of $x$ can be finite
dimensional; in that case it makes sense to count the towers graded
by $x$. We can introduce the character of this space by\begin{equation}
X(q)=\sum_{x}d_{x}q^{x}\label{eq:towercharacter}\end{equation}
where $d_{x}$ is the linear dimension of those towers which have
naive dimension $x$.

The space $\mathcal{L}$ of scaling local boundary operators is also
a graded linear space, where the grading is given by the ultraviolet
scaling weight $\Delta$. The character of that space can be obtained
from boundary conformal field theory. Our aim here is to bring the
two graded spaces into correspondence with each other via comparing
their characters, which is actually nothing else than the counting
of linearly independent local operators classified by their ultraviolet
scaling behaviour.

\section{Scaling Lee-Yang model with boundary}

The scaling Lee-Yang model with boundary is a combined bulk and boundary
perturbation of the boundary version of the $\mathcal{M}_{2,5}$ Virasoro
minimal model which was investigated in detail in \cite{DPTW1}. The
conformal field theory has central charge $c=-22/5$ and the Virasoro
algebra has two irreducible representations $V_{h}$ with highest
weight $h=h_{1,1}=0$ and $h=h_{1,2}=-1/5$ \cite{lycft}. We define
the truncated characters of these representations by\[
\tilde{\chi}_{r,s}(q)=\mathrm{Tr}_{V_{h}}q^{L_{0}-h_{r,s}}=\sum_{n=0}^{\infty}d(n)q^{n}\]
where $d(n)$ gives the degeneracy of the level $n$ descendents.
They can be represented as the following fermionic sums \cite{fermchars}\begin{equation}
\tilde{\chi}_{1,1}(q)=\sum_{n=1}^{\infty}\frac{1}{(1-q^{2+5n})(1-q^{3+5n})}\quad,\quad\tilde{\chi}_{1,2}(q)=\sum_{n=1}^{\infty}\frac{1}{(1-q^{1+5n})(1-q^{4+5n})}\label{eq:trunclychars}\end{equation}

Boundary conformal field theory was developed in \cite{cardybcft,bulkboundaryope,sewing}
and the interested reader is referred to them for details. Applying
the formalism to the conformal Lee-Yang model it can be seen that
there are two conformally invariant boundary conditions. On one of
them, denoted by \textbf{$\mathbf{1}$} in \cite{DPTW1}, the spectrum
of boundary fields is given by the vacuum representation $V_{0}$
of the Virasoro algebra, and therefore it does not have any relevant
boundary fields - thus can have no boundary perturbation either. In
the other case, denoted $\Phi$ in \cite{DPTW1}, the spectrum of
boundary fields is given by the direct sum $V_{0}\oplus V_{-1/5}$
and therefore it has a nontrivial relevant boundary field $\varphi$
with scaling dimension $-1/5$ and the general perturbed boundary
conformal field theory action can be written as%
\footnote{The field $L_{-1}\varphi=\partial_{y}\varphi$, albeit relevant, does
not induce any nontrivial perturbation because it is a total derivative.%
}\[
{\mathcal{A}}_{\lambda,\Phi(h)}={\mathcal{A}}_{\Phi}+\mu\int\limits _{-\infty}^{\infty}dy\int\limits _{-\infty}^{0}dx\phi(x,y)+\mu_{B}\int\limits _{-\infty}^{\infty}dy\varphi(y)\]
where ${\mathcal{A}}_{\Phi}$ denotes the action for ${\mathcal{M}}(2/5)$
with the $\Phi$ boundary condition imposed at $x=0$, and $\mu$
and $\mu_{B}$ denote the bulk and boundary couplings respectively.
There is a unique nontrivial relevant bulk perturbation given by the
spinless field $\phi$ with scaling dimensions $h=\bar{h}=-1/5$.
The action of ${\mathcal{A}}_{\lambda,1}$ is similar, but the last
term on the right hand side is missing. For $\mu>0$ the bulk behaviour
is described by an integrable massive theory having only a single
particle with mass $m$ with the following S matrix \cite{CM}:\[
S(\theta)=-\left(\frac{1}{3}\right)\left(\frac{2}{3}\right)=-\left[\frac{1}{3}\right]\quad;\quad(x)=\frac{\sinh\left(\frac{\theta}{2}+\frac{i\pi x}{2}\right)}{\sinh\left(\frac{\theta}{2}-\frac{i\pi x}{2}\right)}\quad[x]=(x)(1-x)\]
The pole at $\theta=\frac{2\pi i}{3}$ corresponds to the {}``$\varphi^{3}$
property'', i.e. the particle appears as a bound state of itself.
The minimal bulk two-particle form factor only has a zero at $\theta=0$
and a pole at $\theta=\frac{2\pi i}{3}$ in the strip $0\leq\Im m(\theta)<\pi$
and is of the form \cite{Z1}:\begin{equation}
f(\theta)=\frac{y-2}{y+1}v(i\pi-\theta)v(-i\pi+\theta)\quad,\quad y=e^{\theta}+e^{-\theta}\label{eq:min2pffly}\end{equation}
 where\[
v(\theta)=\exp\left\{ 2\int_{0}^{\infty}\frac{dt}{t}e^{i\frac{\theta t}{\pi}}\frac{\sinh\frac{t}{2}\sinh\frac{t}{3}\sinh\frac{t}{6}}{\sinh^{2}t}\right\} \]
and it satisfies\begin{eqnarray*}
f(\theta)f(\theta+i\pi) & = & \frac{\sinh\theta}{\sinh\theta-i\sin\frac{\pi}{3}}\\
f\left(\theta+i\frac{\pi}{3}\right)f\left(\theta-i\frac{\pi}{3}\right) & = & \frac{\cosh\theta+1/2}{\cosh\theta+1}f(\theta)\end{eqnarray*}
and consequently also $f(\theta)\sim O(1)$ for large $\theta$.

In the boundary theory with the perturbed $\Phi$ boundary, the reflection
amplitude of the particle depends on the strength of the coupling
constant of the boundary field as \cite{DPTW1}\[
R(\theta)_{\Phi}=R_{0}(\theta)R(b,\theta)=\left(\frac{1}{2}\right)\left(\frac{1}{6}\right)\left(-\frac{2}{3}\right)\left[\frac{b+1}{6}\right]\left[\frac{b-1}{6}\right]\]
with the one-particle coupling\[
g_{\Phi}=i2(3)^{1/4}(2-\sqrt{3})^{1/2}\frac{\sqrt{3}+2\sin\frac{\pi b}{6}}{\sqrt{3}-2\sin\frac{\pi b}{6}}\]
 where the relation between the dimensionless bootstrap parameter
$b$ and the dimensionful Lagrangian parameters $\mu$ and $\mu_{B}$
is known explicitely \cite{dtwgfunct}, while in the case of the $\mathbf{1}$
boundary the reflection amplitude is the parameter independent expression\[
R(\theta)_{1}=\left(\frac{1}{2}\right)\left(\frac{1}{6}\right)\left(-\frac{2}{3}\right)\]
and the one-particle coupling is\[
g_{1}=-i2(3)^{1/4}(2-\sqrt{3})^{1/2}\]

\subsection{Counting towers for the boundary condition $\mathbf{1}$}

\subsubsection{The recursive equations for the form factors }

The 1PFF function corresponding to the parameter free reflection factor
$R(\theta)_{1}$ is\[
r_{1}(\theta)=i\sinh\theta\,\, u(\theta),\]
 where\[
u(\theta)=\exp\left\{ \int_{0}^{\infty}\frac{dt}{t}\left[\frac{1}{\sinh\frac{t}{2}}-2\cosh\frac{t}{2}\cos\left[\left(\frac{i\pi}{2}-\theta\right)\frac{t}{\pi}\right]\frac{\sinh\frac{5t}{6}+\sinh\frac{t}{2}-\sinh\frac{t}{3}}{\sinh^{2}t}\right]\right\} \]
and asymptotically behaves as $r_{1}\sim\mathrm{e}^{2\theta}$ when
$\theta\rightarrow\infty$. Following the general ideas in 2.2.1 we
introduce the Ansatz\begin{equation}
F_{n}(\theta_{1},\dots,\theta_{n})=4^{n}H_{n}Q_{n}(y_{1},\dots,y_{n})\prod_{i}\frac{r_{1}(\theta_{i})}{y_{i}}\prod_{i<j}\frac{f(\theta_{i}-\theta_{j})f(\theta_{i}+\theta_{j})}{y_{i}+y_{j}}\label{Ansatz1}\end{equation}
where\begin{equation}
H_{n}=\left(\frac{i3^{\frac{1}{4}}}{2^{\frac{1}{2}}v(0)}\right)^{n}\label{eq:Hn}\end{equation}
Then one finds the following recursion equations for the $Q_{n}$-s
\cite{bffprogram}:\begin{eqnarray}
\mathcal{D}: &  & Q_{2}(y_{+},y_{-})=Q_{1}(y)\nonumber \\
 &  & Q_{n+2}(y_{+},y_{-},y_{1},\dots,y_{n})=\prod_{i=1}^{n}(y+y_{i})\, Q_{n+1}(y,y_{1},\dots,y_{n})\quad,\quad n>0;\nonumber \\
\mathcal{K}: &  & Q_{2}(-y,y)=0\nonumber \\
 &  & Q_{n+2}(-y,y,y_{1},\dots,y_{n})=P_{n}(y|y_{1},\dots y_{n})\, Q_{n}(y_{1},\dots,y_{n})\quad,\quad n>0;\nonumber \\
\mathcal{B}: &  & Q_{1}(y)=0\nonumber \\
 &  & Q_{n+1}(0,y_{1},\dots y_{n})=B_{n}(y_{1},\dots,y_{n})\, Q_{n}(y_{1},\dots,y_{n})\quad,\quad n>0;\label{eq:ylrecrel1}\end{eqnarray}
where\begin{equation}
P_{n}(y|y_{1},\dots y_{n})=\frac{1}{2(y_{+}-y_{-})}\left[\prod_{i=1}^{n}(y_{i}-y_{-})(y_{i}+y_{+})-\prod_{i=1}^{n}(y_{i}+y_{-})(y_{i}-y_{+})\right]\label{eq:Pequ}\end{equation}
\begin{equation}
B_{n}(y_{1},\dots,y_{n})=\frac{1}{2\sqrt{3}}\left(\prod_{i=1}^{n}(y_{i}+\sqrt{3})-\prod_{i=1}^{n}(y_{i}-\sqrt{3})\right)\label{eq:Bequ}\end{equation}
and\begin{eqnarray}
y_{+} & = & \omega z+\omega^{-1}z^{-1}\label{eq:ypmeq}\\
y_{-} & = & \omega^{-1}z+\omega z^{-1}\qquad,\qquad\omega=\mathrm{e}^{\frac{i\pi}{3}}\nonumber \end{eqnarray}
with the auxiliary variable $z$ defined as a solution of $y=z+z^{-1}$
(i.e. writing $y=2\cosh\theta$ we obtain $z=\mathrm{e}^{\theta}$).
The symmetry of the expressions (\ref{eq:ylrecrel1}, \ref{eq:Pequ})
in $y_{\pm}$ ensures that the resulting relations do not depend on
which of the two possible solutions of the relation $y=z+z^{-1}$
is chosen (switching from $z=\mathrm{e}^{\theta}$ to $z=\mathrm{e}^{-\theta}$
only interchanges $y_{+}$ with $y_{-}$).

Note also that $Q_{n+2}(y_{+},y_{-},y_{1},\dots,y_{n})$ only depends
on $y$ rather than separately on $y_{\pm}$ because due to its symmetry
in all variables it can only depend on the combinations $y_{+}+y_{-}=y$
and $y_{+}y_{-}=y^{2}-3$. Furthermore it is easy to see that $P_{n}$
is a polynomial in all of its variables (since the expression inside
the parentheses vanishes if $y_{+}=y_{-}$ and is therefore divisible
by $y_{+}-y_{-}$), and using the previous argument again it only
depends on $y$. However, elimination of $y_{\pm}$ makes the formulae
rather cumbersome and therefore we prefer to keep them in our equations. 

It is also interesting to note that the boundary kinematical recursion
is actually {}``almost'' entirely contained in the bulk recursion.
Using the bulk kinematical recursion with $y=0$ results in\[
Q_{n+2}(0,0,y_{1},\dots,y_{n})=P_{n}(0|y_{1}\dots y_{n})Q_{n}(y_{1},\dots,y_{n})\]
while using the boundary kinematical recursion twice leads to\[
Q_{n+2}(0,0,y_{1},\dots,y_{n})=B_{n}(0,y_{1}\dots y_{n})B_{n}(y_{1}\dots y_{n})Q_{n}(y_{1},\dots,y_{n})\]
and therefore it must be true that\begin{equation}
P_{n}(0|y_{1}\dots y_{n})=B_{n}(0,y_{1}\dots y_{n})B_{n}(y_{1}\dots y_{n})\label{eq:PBrelation}\end{equation}
which is indeed satisfied by the expression (\ref{eq:Pequ}, \ref{eq:Bequ}).
This means that the {}``square'' of the boundary kinematical recursion
is contained in the bulk one, as already discussed in \cite{bffprogram}.
The only independent information that the boundary kinematical singularity
axiom carries is the sign of the one-particle coupling $g$: all the
other axioms depend only on the one-particle reflection factor $R$,
which in turn contains only $g^{2}$ using eqn. (\ref{eq:gdef}).

Indeed it was shown by Dorey, Tateo and Watts in \cite{DTWbct} that
although the fundamental reflection factors of the boundary condition
$\mathbf{1}$ and $\Phi(b=0)$ are identical, these boundary conditions
are physically different. Their one-particle couplings differ by a
sign\[
g_{1}=-i2(3)^{1/4}(2-\sqrt{3})^{1/2}=-g_{\Phi(b=0)}\]
and, as a result the whole bootstrap structure changes: the $\mathbf{1}$
boundary condition has no boundary excited state, while the $\Phi(b=0)$
does. This is also reflected in the different expression for their
1PFF function $r$ as discussed after eqn. (\ref{eq:1pffphi}). We
note that while the sign of $g$ is not manifest in the reflection
factor itself, it affects many physical quantities besides the spectrum
such as finite size corrections to the energy levels \cite{DPTW1,BLusch}
and one-point functions of the bulk fields \cite{DPTW1pt,bstate}.

\subsubsection{Asymptotic behaviour of form factors and simple towers}

It can be easily deduced that the asymptotic behaviour of the Ansatz
(\ref{Ansatz1}) is given by\[
\left|F_{n}(\theta_{1}+\theta,\dots,\theta_{n}+\theta)\right|\mathop{\sim}_{\theta\,\rightarrow\,+\infty}\mathrm{e}^{x_{n}\left|\theta\right|}\quad\mathrm{where}\quad x_{n}=\mathrm{deg}\, Q_{n}-\frac{n(n-3)}{2}\]
where $\mathrm{deg}\, P$ denotes the total degree of the multivariable
polynomial $P$ defined by\[
P(\lambda y_{1},\dots,\lambda y_{n})\sim\lambda^{\mathrm{deg}\, P}\quad\mathrm{as}\quad\lambda\rightarrow\infty\]
Note that \[
\mathrm{deg}\,\mathcal{D}(Q_{n+1})=\mathrm{deg}\,\mathcal{B}(Q_{n+1})=\mathrm{deg}\, Q_{n+1}-n+1\;,\;\mathrm{deg}\,\mathcal{K}(Q_{n+2})=\mathrm{deg}\, Q_{n+2}-2n+1\]
As a result, the solutions of the recursion relation can be expanded
in the basis of \emph{simple towers} which are defined to be towers
with the property\begin{eqnarray*}
Q_{n}=0 & : & n<n_{0}\\
\mathrm{deg}\, Q_{n+1}=\mathrm{deg}\, Q_{n}+n-1 & : & n\geq n_{0}\end{eqnarray*}
for some $n_{0}\in\mathbb{N}$. 

For a simple tower the exponent $x_{n}$ describing the asymptotic
growth of the form factor tower is independent of the level $n$.
Therefore we can use the naive dimension to introduce a grading on
the space of towers: the subspace of towers of grade $x$ is spanned
by the simple towers with naive scaling dimension $x$. For a simple
tower given by the polynomials $\left\{ Q_{n}\right\} _{n\in\mathbb{N}}$
the scaling dimension can be expressed as\[
x=\mathrm{deg}\, Q_{n}-\frac{n(n-3)}{2}\]

\subsubsection{The recursion kernel and the number of independent towers}

Here we carry over an approach developed by Koubek \cite{koubek1,koubek2}
to count solutions of the \emph{bulk} form factor equations to the
boundary case. The idea is to classify the so-called 'kernels' of
the recursion relations (\ref{eq:genrecrel}) first and use these
to get the dimensions of the spaces of simple towers.

The kernel can be defined as polynomials (at a given level $n$) that
are taken to zero by the appropriate recursion relation in (\ref{eq:genrecrel}).
The kernel of each recursion at level $n$ can be generated by multiplying
an elementary kernel polynomial with an arbitrary symmetric polynomial
of $n$ variables. The generating polynomials are\begin{eqnarray*}
\mathcal{K} & : & K_{n}^{K}(y_{1},\dots,y_{n})=\prod_{1\leq i<j\leq n}(y_{i}+y_{j})\\
\mathcal{D} & : & K_{n}^{B}(y_{1},\dots,y_{n})=\prod_{1\leq i<j\leq n}(y_{i}^{2}+y_{i}y_{j}+y_{j}^{2}-3)\\
\mathcal{B} & : & K_{n}^{B}(y_{1},\dots,y_{n})=\prod_{i=1}^{n}y_{i}\end{eqnarray*}
The common kernel of the three recursions at level $n$ is generated
by the least common multiple of these polynomials which is just equal
to their product:\[
K_{n}(y_{1},\dots,y_{n})=K_{n}^{K}(y_{1},\dots,y_{n})K_{n}^{B}(y_{1},\dots,y_{n})K_{n}^{D}(y_{1},\dots,y_{n})\]
with degree\[
\mathrm{deg}\, K_{n}=\frac{n(3n-1)}{2}\]
and a basis of the kernel polynomials at level $n$ is given by\[
\sigma_{k_{1}}^{(n)}\dots\sigma_{k_{l}}^{(n)}K_{n}\quad,\quad0<k_{1}\leq k_{2}\dots\leq k_{l}\leq n\]
where $\sigma_{k}^{(n)}$ are the elementary symmetric polynomials
of $n$ variables and degree $k$ defined by the generating relation\[
\prod_{i=1}^{n}(z+y_{i})=\sum_{k=0}^{n}z^{n-k}\sigma_{k}^{(n)}(y_{1},\dots,y_{n})\]
The number of linearly independent towers can be counted using the
observation that every independent tower starts from a kernel polynomial,
because each new tower appears at some level (number of particles)
$n$ as an ambiguity of the solution of the recursion relations. This
is true even for the towers starting at the lowest level $n=1$ if
we note that the kernel at level $1$ is generated by \[
K_{1}(y_{1})=y_{1}\]
The naive scaling dimension of the simple tower starting from the
polynomial $\sigma_{k_{1}}^{(n)}\dots\sigma_{k_{l}}^{(n)}K_{n}$ is\[
x_{k_{1}\dots k_{l}}^{(n)}=\mathrm{deg}\,\sigma_{k_{1}}^{(n)}\dots\sigma_{k_{l}}^{(n)}K_{n}-\frac{n(n-3)}{2}=n(n+1)+k_{1}+\dots+k_{l}\]
Now we can write down the generating function $X(q)$ defined in (\ref{eq:towercharacter})
easily:\[
X(q)=\sum_{n=1}^{\infty}\sum_{m=0}^{\infty}\wp(m|n)q^{n(n+1)+m}\]
where $\wp(m|n)$ denotes the number of partitions of the number $m$
such that none of the summands is greater than $n$. Using the identity\[
\sum_{m=0}^{\infty}\wp(m|n)q^{m}=\prod_{i=1}^{n}\frac{1}{1-q^{i}}\]
we obtain\[
X(q)=\sum_{n=1}^{\infty}\frac{q^{n(n+1)}}{{\displaystyle \prod_{i=1}^{n}1-q^{i}}}\]
and with the help of a Rogers-Ramanujan identity (following \cite{koubek2})\[
X(q)=\sum_{n=1}^{\infty}\frac{1}{(1-q^{2+5n})(1-q^{3+5n})}-1=\tilde{\chi}_{1,1}(q)-1\]
using (\ref{eq:trunclychars}). Therefore the result of the counting
of the form factor towers, graded by their naive scaling dimension,
exactly matches the conformal spectrum of the operators localized
on the $\mathbf{1}$ boundary, which is given by the vacuum representation
of the Virasoro algebra (with highest weight $h_{1,1}=0$). The subtraction
$-1$ corresponds to the identity operator which only has trivial
(vanishing) form factors. 

Note that in claiming the agreement we tacitly supposed that the naive
scaling dimension $x$ of the towers can be identified directly with
the conformal weights of the scaling operators, i.e. that there are
no anomalous dimensions. This was checked for the form factor tower
with the lowest possible dimension $x=2$ in \cite{bffprogram}, where
performing the spectral sum we obtained that the ultraviolet dimension
of the corresponding operator was indeed $\Delta=2$.

\subsubsection{Some explicit solutions}

Finally we give the explicit forms of the lowest levels of some form
factor towers. The $n=1$ generating kernel polynomial%
\footnote{We omit the upper index $n$ of $\sigma_{k}^{(n)}$ specifying the
number of variables since it is always the same as the number of particles,
i.e. the level of the form factor tower.%
} is $K_{1}=\sigma_{1}$, and all kernel polynomials can simply be
written as $\sigma_{1}^{n}$. The first three towers starting at level
$1$ are\begin{eqnarray*}
x=2 & : & Q_{1}^{T}=\sigma_{1}\;,\; Q_{2}^{T}=\sigma_{1}\;,\; Q_{3}^{T}=\sigma_{1}^{2}\;,\; Q_{4}^{T}=\sigma_{1}^{2}(\sigma_{2}+3)\;\dots\\
x=3 & : & Q_{n}^{\partial T}=\sigma_{1}Q_{n}^{T}\\
x=4 & : & Q_{n}^{\partial^{2}T}=\sigma_{1}^{2}Q_{n}^{T}\end{eqnarray*}
The upper index $T$ shows that the tower corresponds to the (off-critical
version of the) boundary stress-energy operator $T=L_{-2}\mathbb{I}$
(already found in \cite{bffprogram}), while the other towers are
its first and second derivatives respectively. In fact, there is a
single tower for each value of $x$ less than $6$ which corresponds
to the fact that up to this level (due to the presence of null-vectors)
the conformal vacuum representation of the $c=-22/5$ Virasoro algebra
contains only the vectors $L_{-1}^{n}L_{-2}\mathbb{I}$ ($n=0,\,1,\,2,\,3$).
At $x=6$ the conformal representation contains another linearly independent
vector, and indeed a new tower appears, which starts from the level
$2$ generating kernel polynomial:\begin{eqnarray*}
Q_{2}^{(6)} & = & \sigma_{1}^{2}\sigma_{2}\left(\sigma_{1}^{2}-\sigma_{2}-3\right)\\
Q_{3}^{(6)} & = & \sigma_{1}\left(\sigma_{1}\sigma_{2}-\sigma_{3}\right)\left(\sigma_{1}^{2}-\sigma_{2}-3\right)\\
Q_{4}^{(6)} & = & \sigma_{1}\left(\sigma_{2}+3\right)\left(\left(\sigma_{1}\sigma_{2}-\sigma_{3}\right)\left(\sigma_{1}^{2}-\sigma_{2}-3\right)+\sigma_{1}\sigma_{4}\right)\end{eqnarray*}
The counting argument in the last subsection ensures that the new
towers always appear in such a way that the space of simple towers
graded by $x$ matches the space of linearly independent vectors in
the vacuum representation.

\subsection{Counting towers for the boundary condition $\Phi$}

In this case the 1PFF is \cite{bffprogram}\begin{equation}
r_{\Phi}(\theta)=\frac{i\sinh\theta}{(\sinh\theta-i\sin\frac{\pi(b+1)}{6})(\sinh\theta-i\sin\frac{\pi(b-1)}{6})}\,\, u(\theta)\label{eq:1pffphi}\end{equation}
The main difference from the $\mathbb{I}$ case is the presence of
the denominators, which correspond to boundary excited states. Note
that at $b=0$ the reflection factor $R(\theta)_{\Phi}$ is identical
to $R(\theta)_{1}$, but the corresponding 1PFFs are different. This
is related to the different interpretation of the pole structure of
the $R$ matrices discussed in detail in \cite{DTWbct}: for the boundary
condition $\mathbf{1}$ the boundary excited state is absent and the
corresponding pole at $\theta=i\frac{\pi}{6}$ is explained by a Coleman-Thun
diagram. In this case we take the following Ansatz\begin{equation}
F_{n}(\theta_{1},\dots,\theta_{n})=H_{n}Q_{n}(y_{1},\dots,y_{n})\prod_{i}\frac{r_{\Phi}(\theta_{i})}{y_{i}}\prod_{i<j}\frac{f(\theta_{i}-\theta_{j})f(\theta_{i}+\theta_{j})}{y_{i}+y_{j}}\label{AnsatzPhi}\end{equation}
The recursion relations in this case are \cite{bffprogram}\begin{eqnarray*}
\mathcal{D}: &  & Q_{2}(y_{+},y_{-})=(y^{2}-\beta_{-3}^{2})Q_{1}(y)\\
 &  & Q_{n+2}(y_{+},y_{-},y_{1},\dots,y_{n})=(y^{2}-\beta_{-3}^{2})\prod_{i=1}^{n}(y+y_{i})\, Q_{n+1}(y,y_{1},\dots,y_{n})\quad,\quad n>0;\\
\mathcal{K}: &  & Q_{2}(-y,y)=0\\
 &  & Q_{n+2}(-y,y,y_{1},\dots,y_{n})=(y^{2}-\beta_{-1}^{2})(y^{2}-\beta_{1}^{2})P_{n}(y|y_{1},\dots y_{n})Q_{n}(y_{1},\dots,y_{n})\quad,\quad n>0;\\
\mathcal{B}: &  & Q_{1}(0)=0\\
 &  & Q_{n+1}(0,y_{1},\dots y_{n})=\beta_{-1}\beta_{1}B_{n}(y_{1},\dots y_{n})Q_{n}(y_{1},\dots,y_{n})\quad,\quad n>0;\end{eqnarray*}
where $P_{n}$ and $B_{n}$ are given by eqns. (\ref{eq:Pequ}, \ref{eq:Bequ},
\ref{eq:ypmeq}) and\[
\beta_{k}(b)=2\cos\frac{\pi}{6}(b+k)\]
The bulk and boundary kinematical recursions again satisfy the appropriate
generalization of the compatibility relation (\ref{eq:PBrelation}).

For the degrees of the recursions we have\[
\mathrm{deg}\,\mathcal{D}(Q_{n+1})=\mathrm{deg}\,\mathcal{B}(Q_{n+1})=\mathrm{deg}\, Q_{n+1}-n-1\;,\;\mathrm{deg}\,\mathcal{K}(Q_{n+2})=\mathrm{deg}\, Q_{n+2}-2n-3\]
and so simple towers can now be defined by the property\begin{eqnarray*}
Q_{n}=0 & : & n<n_{0}\\
\mathrm{deg}\, Q_{n+1}=\mathrm{deg}\, Q_{n}+n+1 & : & n\geq n_{0}\end{eqnarray*}
Due to the additional factors in the denominator the asymptotic behaviour
of $r_{\Phi}(\theta)$ for $\theta\rightarrow\infty$ is $O(1)$,
and so the naive scaling dimension of a simple tower corresponding
to the polynomials $\left\{ Q_{n}\right\} _{n\in\mathbb{N}}$ is\[
x=\mathrm{deg}\, Q_{n}-\frac{n(n+1)}{2}\]
The kernel is generated by the same polynomials as before, but now
the naive scaling dimension corresponding to $\sigma_{k_{1}}^{(n)}\dots\sigma_{k_{l}}^{(n)}K_{n}$
is\[
x_{k_{1}\dots k_{l}}^{(n)}=\mathrm{deg}\,\sigma_{k_{1}}^{(n)}\dots\sigma_{k_{l}}^{(n)}K_{n}-\frac{n(n+1)}{2}=n(n-1)+k_{1}+\dots+k_{l}\]
and as a result we obtain\begin{eqnarray*}
X(q) & = & \sum_{n=1}^{\infty}\sum_{m=0}^{\infty}\wp(m|n)q^{n(n-1)+m}=\sum_{n=1}^{\infty}\frac{q^{n(n-1)}}{{\displaystyle \prod_{i=1}^{n}1-q^{i}}}=\sum_{n=1}^{\infty}\frac{q^{n(n+1)}}{{\displaystyle \prod_{i=1}^{n}1-q^{i}}}+\sum_{n=1}^{\infty}\frac{q^{n^{2}}}{{\displaystyle \prod_{i=1}^{n}1-q^{i}}}\\
 & = & \sum_{n=1}^{\infty}\frac{1}{(1-q^{2+5n})(1-q^{3+5n})}+\sum_{n=1}^{\infty}\frac{1}{(1-q^{1+5n})(1-q^{4+5n})}-1\\
 & = & \tilde{\chi}_{1,1}(q)+\tilde{\chi}_{1,2}(q)-1\end{eqnarray*}
using again a Rogers-Ramanujan identity following \cite{koubek2}.
This is consistent with the fact that in the conformal limit the boundary
operator content of the $\Phi$ boundary condition is given by the
direct sum of the vacuum module with highest weight $h_{1,1}=0$ and
the only other nontrivial representation with highest weight $h_{1,2}=-1/5$.

Note that in this case to achieve the agreement we must suppose that
the naive scaling dimensions of the towers corresponding to operators
of the identity representation do not get any anomalous corrections,
while those corresponding to operators in the model with highest weight
$h_{1,2}=-1/5$ get exactly the right contribution for their conformal
weight to match with that predicted by conformal field theory. The
true scaling weight of the operator can only be computed by examining
appropriate two-point functions. This makes the operator identification
very difficult for higher levels as shown below using the example
of the boundary stress-energy tensor.

For $x=0$ the operator identification is easy, since there is a unique
tower with the required asymptotic properties: \begin{eqnarray}
Q_{1}^{\varphi} & = & \sigma_{1}\,,\quad Q_{2}^{\varphi}=\sigma_{1}(\sigma_{2}+3-\beta_{-3}^{2})\,,\quad\nonumber \\
Q_{3}^{\varphi} & = & \sigma_{1}\left[\sigma_{1}(\sigma_{2}+\beta_{-1}^{2})(\sigma_{2}+\beta_{1}^{2})-(\sigma_{2}+3)(\sigma_{1}\sigma_{2}-\sigma_{3})\right]\,,\dots\label{eq:fitower}\end{eqnarray}
 In \cite{bffprogram} the two-point function of this tower was compared
to the perturbed conformal field theory prediction and we found that
its identification with the operator $\varphi$ is justified (this
also settles the case for all the derivatives of $\varphi$). We remark
that the anomalous dimension of the correlator in this case turns
out to be\[
2\Delta=-\frac{1}{5}\neq2h=-\frac{2}{5}\]
which is due to the fact that $\varphi$ itself appears in the $\varphi\varphi$
operator product, and so the above result is in accordance with the
discussion under 2.2.2 since $h_{\mathrm{min}}=-1/5$ (cf. also \cite{bffprogram}
and also the analogous bulk situation discussed by Zamolodchikov in
\cite{Z1}). For $x=1$ there is again a single operator\[
Q_{n}^{\partial\varphi}=\sigma_{1}Q_{n}^{\varphi}\]
whose scaling dimension can be known exactly in terms of the scaling
dimensions of the lowest tower, since it is a derivative operator.
However, we find two towers at $x=2$. One is the second derivative\begin{equation}
Q_{n}^{\partial^{2}\varphi}=\sigma_{1}^{2}Q_{n}^{\varphi}\label{eq:d2phi}\end{equation}
while the other one starts from the $n=2$ generating kernel polynomial:\[
Q_{2}^{\tilde{T}}=\sigma_{1}\sigma_{2}\left(\sigma_{1}^{2}-\sigma_{2}-3\right)\,,\quad Q_{3}^{\tilde{T}}=\sigma_{1}\left(\sigma_{1}\sigma_{2}-\sigma_{3}\right)\left(\beta_{1}\beta_{-1}\left(\sigma_{1}^{2}-\sigma_{2}-3\right)+\sigma_{1}\sigma_{3}\right)\,,\dots\]
which is consistent with the fact that both the Virasoro module $h=\-1/5$
and the module $h=0$ have a single linearly independent vector at
level two: $L_{-1}^{2}\varphi$ and $T=L_{-2}\mathbb{I}$, respectively.
The tower (\ref{eq:d2phi}) can be directly identified with $L_{-1}^{2}\varphi$
since the action of $L_{-1}$ is exactly given by the derivative,
but the identification of $T$ is more involved. In general we expect
that\begin{equation}
T\propto\tilde{T}+\alpha\varphi+\beta\partial\varphi+\gamma\partial^{2}\varphi\label{eq:Tmixing}\end{equation}
(where the proportionality sign means that the normalization of the
operator must also be fixed). Note that $T$ cannot mix with higher
towers because that would spoil the short-distance behaviour. The
identification of $T$ is an open question, to which we return in
the conclusions.

\section{Sinh-Gordon model with Dirichlet boundary conditions}

The sinh-Gordon theory in the bulk is defined by the Lagrangian\[
\mathcal{L}=\frac{1}{2}(\partial_{\mu}\Phi)^{2}-\frac{m^{2}}{b^{2}}(\cosh b\Phi-1)\]
It can be considered as the analytic continuation of the sine-Gordon
model for imaginary coupling $\beta=ib$. The S-matrix of the model
is\[
S(\theta)=-\left(1+\frac{B}{2}\right)\left(-\frac{B}{2}\right)=-\left[-\frac{B}{2}\right]\qquad;\quad B=\frac{2b^{2}}{8\pi+b^{2}}\]
In this case there is no self-fusion pole, and so bulk dynamical singularities
are absent. The minimal bulk two-particle form factor belonging to
this S-matrix is \cite{FMS}\[
f(\theta)=\mathcal{N}\exp\left[8\int_{0}^{\infty}\frac{dx}{x}\sin^{2}\left(\frac{x(i\pi-\theta)}{2\pi}\right)\frac{\sinh\frac{xB}{4}\sinh(1-\frac{B}{2})\frac{x}{2}\sinh\frac{x}{2}}{\sinh^{2}x}\right],\]
 satisfying\begin{equation}
f(\theta)f(\theta+i\pi)=\frac{\sinh\theta}{\sinh\theta+i\sin\frac{\pi B}{2}}\label{eq:ffip}\end{equation}
and therefore $f(\theta)\sim O(1)$ as $\theta\rightarrow\infty$. 

Sinh-Gordon theory can be restricted to the negative half-line, but
integrability is only maintained by imposing either the Dirichlet\[
\Phi(0,t)=\Phi_{0}^{D}\]
or the two parameter family of perturbed Neumann\[
V_{B}(\Phi(0,t))=M_{0}\cosh\left(\frac{b}{2}(\Phi(0,t)-\Phi_{0})\right)-M_{0}\]
boundary conditions \cite{GZ}. The latter interpolates between the
Neumann and the Dirichlet boundary conditions, since for $M_{0}=0$
we recover the Neumann, while for $M_{0}\to\infty$ the Dirichlet
boundary condition with $\Phi_{0}^{D}=\Phi_{0}$. The reflection factor
which depends on two continuous parameters can be written as \cite{shGEF}\[
R(\theta)=R_{0}(\theta)R(E,F,\theta)=\left(\frac{1}{2}\right)\left(\frac{1}{2}+\frac{B}{4}\right)\left(-1-\frac{B}{4}\right)\left[\frac{E-1}{2}\right]\left[\frac{F-1}{2}\right]\]
 and can be obtained as the analytic continuation of the first breather
reflection factor in boundary sine-Gordon model which was calculated
by Ghoshal in \cite{ghoshal}. The relation of the bootstrap parameters
$E$ and $F$ to the parameters of the Lagrangian is known both from
a semiclassical calculation \cite{shGEF,corrigan} and also in an
exact form in the perturbed boundary conformal field theory framework
\cite{sinG_uvir}. Imposing Dirichlet boundary condition instead of
the general one corresponds to removing the $F$ dependent factor
from $R(\theta)$.

\subsection{Recursion relations for the form factors with Dirichlet boundary
conditions}

Boundary form factors for sinh-Gordon theory with Dirichlet boundary
condition have already been investigated to some extent in \cite{bffprogram},
and many solutions were constructed by Castro-Alvaredo \cite{alvaredo}
at the self-dual point $B=1$. Here we aim to classify all the solutions
for general coupling.

The 1PFF corresponding to Dirichlet boundary condition is \cite{bffprogram}\[
r_{E}(\theta)=\frac{\sinh\theta}{\sinh\theta-i\sin\gamma}u(\theta,B)\]
where\[
u(\theta,B)=\exp\left[-2\int_{0}^{\infty}\frac{dx}{x}\left[\cos(\frac{i\pi}{2}-\theta)\frac{x}{\pi}-1\right]\frac{\cosh\frac{x}{2}}{\sinh^{2}x}\left(\sinh\frac{xB}{4}+\sinh(1-\frac{B}{2})\frac{x}{2}+\sinh\frac{x}{2}\right)\right]\]
and\[
\gamma=\frac{\pi}{2}(E-1)\]
The asymptotics of the 1PFF is $r_{E}\sim\mathrm{e}^{\theta}$ as
$\theta\rightarrow\infty$. The one-particle coupling to the boundary
is given by\[
g_{E}=\frac{2(1+\cos\frac{\pi B}{4}+\sin\frac{\pi B}{4})}{\sqrt{\sin\frac{\pi B}{2}}}\frac{\cos\gamma}{1-\sin\gamma}\]
Writing the $n$-particle form factors in the general form\[
F_{n}(\theta_{1},\dots,\theta_{n})=H_{n}Q_{n}(y_{1},\dots,y_{n})\prod_{i}\frac{r_{E}(\theta_{i})}{y_{i}}\prod_{i<j}\frac{f(\theta_{i}-\theta_{j})f(\theta_{i}+\theta_{j})}{y_{i}+y_{j}},\]
and choosing the ratio of the $H_{n}$-s appropriately the recursion
equations take the form:\begin{eqnarray}
\mathcal{K}: &  & Q_{2}(-y,y)=0\nonumber \\
 &  & Q_{n+2}(-y,y,y_{1},\dots,y_{n})=(y^{2}-4\cos^{2}\gamma)P_{n}(y|y_{1},\dots,y_{n})\, Q_{n}(y_{1},\dots,y_{n})\quad,\quad n>0;\nonumber \\
\mathcal{B}: &  & Q_{1}(0)=0\nonumber \\
 &  & Q_{n+1}(0,y_{1},\dots,y_{n})=2\cos\gamma B_{n}(y_{1},\dots,y_{n})\, Q_{n}(y_{1},\dots,y_{n})\quad,\quad n>0;\label{eq:sinhgrecursions}\end{eqnarray}
where $P_{n}$ is given by eqns. (\ref{eq:Pequ}, \ref{eq:ypmeq})
with $\omega=e^{-i\pi\frac{B}{2}}$,\[
B_{n}(y_{1},\dots,y_{n})=\frac{1}{4\sin\frac{\pi B}{2}}\left(\prod_{i=1}^{n}\left(y_{i}-2\sin\frac{\pi B}{2}\right)-\prod_{i=1}^{n}\left(y_{i}+2\sin\frac{\pi B}{2}\right)\right)\]
and the bulk and boundary kinematical recursions again satisfy the
appropriate generalization of the compatibility relation (\ref{eq:PBrelation}).
Note that for $E=0$ the right-hand side of the boundary kinematical
relation is zero:\[
\mathcal{B}:\; Q_{n+1}(0,y_{1},\dots,y_{n})=0\]
and so one could dispense with the boundary kinematical recursion
choosing an Ansatz of the form\[
F_{n}(\theta_{1},\dots,\theta_{n})=H_{n}Q_{n}(y_{1},\dots,y_{n})\prod_{i}r_{E=0}(\theta_{i})\prod_{i<j}\frac{f(\theta_{i}-\theta_{j})f(\theta_{i}+\theta_{j})}{y_{i}+y_{j}}\]
as in \cite{bffprogram}. In this paper, however, we choose to treat
the two cases together.

\subsection{Counting towers and the spectrum of local operators}

The recursion relations (\ref{eq:sinhgrecursions}) have the following
degrees:\[
\mathrm{deg}\,\mathcal{K}(Q_{n+2})=\mathrm{deg}\, Q_{n+2}-2n-1\;,\;\mathrm{deg}\,\mathcal{B}(Q_{n+1})=\mathrm{deg}\, Q_{n+1}-n\]
and so simple towers are defined as\begin{eqnarray*}
Q_{n}=0 & : & n<n_{0}\\
\mathrm{deg}\, Q_{n+1}=\mathrm{deg}\, Q_{n}+n & : & n\geq n_{0}\end{eqnarray*}
while the naive scaling dimension of a simple tower corresponding
to the polynomials $\left\{ Q_{n}\right\} _{n\in\mathbb{N}}$ is given
by\[
x=\mathrm{deg}\, Q_{n}-\frac{n(n-1)}{2}\]
Due to the absence of bulk dynamical singularities, the generating
kernel polynomial at level $n$ is the product of the ones for the
bulk and boundary kinematical recursions\[
K_{n}'(y_{1},\dots,y_{n})=K_{n}^{K}(y_{1},\dots,y_{n})K_{n}^{B}(y_{1},\dots,y_{n})\]
with degree\[
\mathrm{deg}\, K_{n}'=\frac{n(n+1)}{2}\]
and the naive scaling dimension of the simple tower starting from
the polynomial $\sigma_{k_{1}}^{(n)}\dots\sigma_{k_{l}}^{(n)}K_{n}'$
is\[
x_{k_{1}\dots k_{l}}^{(n)}=\mathrm{deg}\,\sigma_{k_{1}}^{(n)}\dots\sigma_{k_{l}}^{(n)}K_{n}'-\frac{n(n-1)}{2}=n+k_{1}+\dots+k_{l}\]
The generating function (\ref{eq:towercharacter}) takes the form\[
X(q)=\sum_{n=1}^{\infty}\sum_{m=0}^{\infty}\wp(m|n)q^{n+m}=\sum_{n=1}^{\infty}\wp(n)q^{n}=\prod_{j=1}^{\infty}\frac{1}{1-q^{j}}\]
where $\wp(n)$ is simply the number of all partitions of $n$.

On the other hand, the independent boundary operators for the Dirichlet
boundary condition $\Phi(t,x=0)=\Phi_{0}$ are simply given by the
differential monomials\[
\left.\partial_{x}^{k_{1}}\Phi(t,x)\dots\partial_{x}^{k_{l}}\Phi(t,x)\right|_{x=0}\]
which have scaling weight $k_{1}+\dots+k_{l}$. Therefore the number
of operators with a given scaling weight $n\in\mathbb{N}$ is indeed
equal to the number of integer partitions of $n$.

\subsection{General boundary conditions}

When both $E$ and $F$ are nonzero, the 1PFF reads\[
r_{EF}(\theta)=\frac{\sinh\theta}{(\sinh\theta-i\sin\gamma)(\sinh\theta-i\sin\gamma')}u(\theta,B)\quad,\quad\gamma=\frac{\pi}{2}(E-1)\quad\gamma'=\frac{\pi}{2}(F-1)\]
with asymptotics $r_{EF}\sim O(1)$ when $\theta\rightarrow\infty$,
and therefore the scaling dimension of a simple tower is\[
x=\mathrm{deg}\, Q_{n}-\frac{n(n+1)}{2}\]
and so each tower starting from an elementary kernel polynomial $K_{n}'$
has naive scaling dimension $x=0$. This results in infinitely many
towers corresponding to any integer value of the naive scaling dimension!

This is in fact not so surprising if we consider the expected spectrum
of scaling operators. These are of the form\[
\left.\partial_{t}^{k_{1}}\Phi(t,x)\dots\partial_{t}^{k_{l}}\Phi(t,x)\mathrm{e}^{\alpha\Phi(t,x)}\right|_{x=0}\]
(the $x$-derivatives of the field can be expressed with exponential
operators using the boundary condition). These can be thought to be
organized into families of descendents of the exponential operators\[
\mathrm{e}^{\alpha\Phi(t,x=0)}\]
the descendent level given by $k_{1}+\dots+k_{l}$, so the number
of descendents at some level $n$ is the number of partitions of the
integer $n$. However, the exponential operator has naive (classical)
scaling dimension zero: its scaling weight is a fully quantum effect.
Although at first it seems that there exists a continuum of such operators,
recall that they can be expressed in terms of powers of the field:\[
\mathrm{e}^{\alpha\Phi(t,0)}=\sum_{k=1}^{\infty}\frac{\alpha^{k}}{k\,!}\Phi^{k}(t,0)\]
which shows that they depend only on countably many independent operators,
the powers $\Phi^{k}$ (which are not scaling fields in themselves,
their correlation function involves logarithms) so one expects only
a countable infinity of form factor towers with naive scaling dimension
$0$, which is indeed what we found.

In this paper we cannot go further with the identification between
the form factor towers and the local operators because some further
tools are necessary in order to classify the towers of zero naive
scaling dimension arising from the kernel solutions; see the conclusions
in the next section for details. We only wish to note that our results
in the free field limit are identical to those in \cite{bffprogram},
and that the whole situation above is very similar to the case of
the bulk sinh-Gordon model \cite{koubek_mussardo}.

\section{Conclusions and open problems}

In this paper we performed the classification and counting of the
solutions of the boundary form factor bootstrap based on their ultraviolet
behaviour. We showed that in the boundary version of the scaling Lee-Yang
model and in the boundary sinh-Gordon model the results are in perfect
agreement with the expectations based on boundary conformal field
theory in the first case and on the Lagrangian formulation in the
second. This gives an additional and crucial piece of evidence for
the consistency of the bootstrap axioms developed in \cite{bffprogram}.

Our discussion, however, has made several open problems manifest,
that are worthwhile to be pursued in the future.

\subsection{The issue of anomalous dimensions}

The first is the question of anomalous dimensions. It was shown (using
the example of the boundary operator $\varphi$ and its descendents
in the Lee-Yang case) that the naive dimension obtained from the asymptotic
growth of the form factors of some tower is not necessarily identical
to the exact ultraviolet weight of the corresponding operator. In
fact the counting of towers depends very much on the additional assumption
that anomalous contributions to the scaling dimensions do not mess
up the spectrum%
\footnote{It could happen, at least in principle, that the grading of the operators
is changed by the anomalous dimensions, so that the classification
according to conformal descendents does not match the grading provided
by the naive scaling dimension.%
}, and so the space of simple towers with a given value of the naive
scaling dimension $x$ can be brought into correspondence with the
space of local operators of a given value for the exact ultraviolet
weight $h$.

This issue is also well-known in the bulk and there the computation
the ultraviolet weight can be tackled by two different approaches.
One of them uses the cumulant expansion of the logarithm of the two-point
function \cite{smirnovcumulant} (for a very nice discussion see also
\cite{babujiankarowski}). The spectral expansion (\ref{eq:2pt})
can be written in the following form for the Euclidean two-point function\begin{eqnarray*}
\rho(m\tau) & = & \sum_{n=0}^{\infty}\frac{1}{n!}\int_{-\infty}^{\infty}d\theta_{1}\int_{-\infty}^{\infty}d\theta_{2}\dots\int_{-\infty}^{\infty}d\theta_{n}e^{-m\tau\sum_{i}\cosh\theta_{i}}f_{n}\left(\theta_{1},\dots,\theta_{n}\right)\\
 &  & f_{n}=\frac{1}{(4\pi)^{n}}F_{n}F_{n}^{+}\end{eqnarray*}
 where we used that the functions $f_{n}$ are symmetric and even
in all the rapidity variables due to the symmetry properties of the
form factors (note that in a unitary theory they are also positive
semidefinite, but that is not necessary for the following argument).
Supposing that the leading $n=0$ term in the short-distance limit
is a constant and normalizing it to $1$ we can write a similar expansion
for the logarithm of the correlation function\[
\log\rho(m\tau)=\sum_{n=1}^{\infty}\frac{1}{n!}\int_{-\infty}^{\infty}d\theta_{1}\int_{-\infty}^{\infty}d\theta_{2}\dots\int_{-\infty}^{\infty}d\theta_{n}e^{-m\tau\sum_{i}\cosh\theta_{i}}h_{n}\left(\theta_{1},\dots,\theta_{n}\right)\]
where the $h_{n}$ are the cumulants of the functions $f_{n}$ defined
recursively by\begin{eqnarray*}
 &  & f_{1}(\theta_{1})=h_{1}(\theta_{1})\quad,\quad f_{2}(\theta_{1},\theta_{2})=h_{2}(\theta_{1},\theta_{2})+h_{1}(\theta_{1})h_{1}(\theta_{2})\\
 &  & f_{3}(\theta_{1},\theta_{2},\theta_{3})=h_{3}(\theta_{1},\theta_{2},\theta_{3})+h_{1}(\theta_{1})h_{2}(\theta_{2},\theta_{3})+h_{1}(\theta_{2})h_{2}(\theta_{1},\theta_{3})+h_{1}(\theta_{3})h_{2}(\theta_{1},\theta_{2})\\
 &  & +h_{1}(\theta_{1})h_{1}(\theta_{2})h_{1}(\theta_{3})\\
 &  & \dots\end{eqnarray*}
Defining \begin{equation}
\bar{h}_{n}(\theta_{1},\dots,\theta_{n})=\lim_{\lambda\rightarrow\infty}h_{n}(\theta_{1}+\lambda,\dots,\theta_{n}+\lambda)\label{eq:hbarlimit}\end{equation}
it is easy to see that the functions $\bar{h}_{n}$ depend only on
the differences of the rapidities (for spinless operators in the bulk
this is already true for $f_{n}$ and therefore also $h_{n}$ and
in that case $\bar{h}_{n}\equiv h_{n}$). Following the same argument
as in the bulk case we arrive at the representation\begin{equation}
\Delta=\sum_{n=1}^{\infty}\frac{1}{n!}\int_{-\infty}^{\infty}d\theta_{2}\dots\int_{-\infty}^{\infty}d\theta_{n}\bar{h}_{n}\left(0,\theta_{2},\dots,\theta_{n}\right)\label{eq:cumulantexpansion}\end{equation}
for the short-distance exponent defined in (\ref{eq:shortdistance}).

The main problem with this approach is that the limit (\ref{eq:hbarlimit})
exists only for boundary operators for which the naive scaling dimension
is zero, and therefore it is of very limited use (in the bulk it obviously
exists for any spinless operator, but a similar problem appears when
trying to apply this approach to operators with non-vanishing Lorentz
spin). An example to which the cumulant expansion (\ref{eq:cumulantexpansion})
can be applied is given by the operator $\varphi$ in the scaling
Lee-Yang model with $\Phi$ boundary condition, but in that case we
already have a much more detailed comparison with the conformal prediction
directly via the two-point function, which was carried out in \cite{bffprogram}.
Therefore it is an interesting problem to develop some method to extract
the ultraviolet dimension directly using only the spectral densities
$f_{n}$, generalizing the above argument.

The other approach in the bulk is to use a sum rule such as the so-called
$\Delta$--theorem developed in \cite{delta-th}. Unfortunately, the
arguments used in the bulk case are not directly applicable to boundary
theories, because the boundary component of the energy-momentum tensor
does not obey any conservation law in itself, but it remains to be
seen whether there exists some other way of deriving a sum rule. 

A further possibility along this line of thought would be to examine
the operator product of $T$ (the operator that corresponds to the
conformal boundary Virasoro current off criticality) with the boundary
operator $\mathcal{O}$ in question using the spectral representation
for their two-point function, since it is known from boundary conformal
field theory that they have an operator product expansion of the form\[
T(\tau)\mathcal{O}(0)=\frac{h_{\mathcal{O}}\mathcal{O}(0)}{\tau^{2}}+\textrm{less singular terms}\]
but to perform this we would need to identify the form factors corresponding
to $T$; the discussion below just highlights the difficulties related
to operator identification.

\subsection{Operator identification}

It is also an open problem how to identify individual operators with
their corresponding towers. In some cases this is straightforward,
like for the operator with the lowest conformal weight (an excellent
example is again the operator $\varphi$ in the scaling Lee-Yang model
with $\Phi$ boundary condition, or the boundary stress energy tensor
$T$ in the case of the $\mathbf{1}$ boundary condition \cite{bffprogram}).
It is also rather easy to identify form factors of derivative operators
since\[
F^{\partial_{\tau}\mathcal{O}}(\theta_{1},\dots,\theta_{n})\propto\left(\sum_{i=1}^{n}\cosh\theta_{i}\right)F^{\mathcal{O}}(\theta_{1},\dots,\theta_{n})\]
(once the form factors of $\mathcal{O}$ are known). For other operators,
however, the identification is not at all straightforward. Take the
example of the boundary stress energy tensor $T$ for the $\Phi$
boundary condition in the scaling Lee-Yang model (subsection 3.2)
where the new tower at $x=2$, corresponding to the appearance of
$T$, mixes non-trivially with other towers according to (\ref{eq:Tmixing}).
It is not easy to disentangle the mixing and so produce a direct identification.
There are certainly some approaches to try which we leave to future
investigations.

One possibility is to calculate the mutual two-point correlation functions
of the towers $\tilde{T}$, $\varphi$, $\partial\varphi$, $\partial^{2}\varphi$
and compare it numerically to the conformal perturbation theory prediction
for the operators $T$, $\varphi$, $L_{-1}\varphi$, $L_{-1}^{2}\varphi$
(in fact only the correlators involving $\tilde{T}$ are necessary).
Using that, one can then fit the mixing (and normalization) coefficients
numerically.

Another approach to operator identification in general is to use a
boundary extension of the approach developed by Delfino and Niccoli
for the bulk scaling Lee-Yang model in \cite{descendents} who used
the fact that part of the conformal descendents can be generated using
derivation ($L_{-1}$) and the charges which remain conserved off-criticality.
This is very efficient for low levels: it makes possible the precise
identification of towers with the appropriate operators up to descendent
level $7$. It remains to be seen whether this method can be extended
to the boundary case.

In general, however, it is also difficult to identify towers corresponding
to the primary fields. As we have shown in subsection 4.3, this is
a very complicated issue in the boundary sinh-Gordon model with general
boundary conditions, where an infinity of $x=0$ towers must be matched
against the spectrum of exponential fields. In the bulk case form
factors of primary fields satisfy a factorization property \cite{Smirnov,smirnovcumulant,mussardo_simonetti},
for which a general argument was given in \cite{delta-th}. Factorization
makes possible the identification of towers corresponding to primary
fields, and can also be extended to descendents. It remains to be
seen whether such clustering property can be extended to the boundary
form factor bootstrap: at first sight, it seems more promising to
take an approach which relates clustering to the operator product
expansion as in the paper \cite{balog_weisz} by Balog and Weisz instead
of one appealing to conformal holomorphic factorization (as done by
Cardy et al. in \cite{delta-th}), which is broken by the presence
of the boundary condition. In this regard it is also interesting to
examine the relation between the bulk and boundary form factor bootstrap,
to which we now turn.

\subsection{Connection to the bulk form factor bootstrap}

Isolating the leading asymptotic coefficient of the $n$-particle
form factor $F_{n}^{\mathcal{O}}$ in the large rapidity limit\[
F_{n}^{\mathcal{O}}(\theta_{1}+\lambda,\dots,\theta_{n}+\lambda)\mathop{\sim}_{\lambda\rightarrow+\infty}\bar{F}_{n}^{\mathcal{O}}(\theta_{1},\dots,\theta_{n})\mathrm{e}^{x\lambda}+\textrm{subleading terms}\]
it is obvious that $\bar{F}_{n}^{\mathcal{O}}(\theta_{1},\dots,\theta_{n})$
is a function of rapidity differences only, and it is also easy to
verify that it satisfies the bulk form factor axioms as a consequence
of $F_{n}^{\mathcal{O}}$ satisfying the boundary form factor axioms.
Therefore $\bar{F}_{n}^{\mathcal{O}}(\theta_{1},\dots,\theta_{n})$
can identified with the $n$-particle form factor $f_{n}^{\tilde{\mathcal{O}}}(\theta_{1},\dots,\theta_{n})$
of some bulk operator $\tilde{\mathcal{O}}$. As an example, using
the asymptotic behaviours of the 2PFF function $f$ in (\ref{eq:min2pffly})
and of the 1PFF function $r_{\Phi}$ in (\ref{eq:1pffphi}), the form
factor tower $F^{\varphi}$ (\ref{eq:fitower}) in the Lee-Yang model
with $\Phi$ boundary condition in this limit gives rise (up to normalization)
to the bulk form factor tower of the trace of stress energy tensor
$\Theta$ found by Zamolodchikov \cite{Z1}. As an application of
this correspondence it is easy to see that the bulk cumulant expansion
for the anomalous dimension of $\Theta$ in \cite{smirnovcumulant}
is just twice the expression (\ref{eq:cumulantexpansion}) evaluated
for the boundary field $\varphi$. This is consistent with the fact
that in the bulk $\Delta_{\Theta}=-2/5$ while for the boundary case
$\Delta_{\varphi}=-1/5$. 

This correspondence may be related in some way to the bulk-boundary
operator product expansion in boundary conformal field theory \cite{bulkboundaryope},
and it is possible that (extending the example above) the exact scaling
dimension of operators can be identified this way using known results
for the bulk form factors. It could also explain why the detailed
structure of the counting of form factor towers (kernels, naive scaling
dimensions, character identities to use etc.) is so similar to the
bulk case as it was treated in \cite{koubek_mussardo,koubek1,koubek2}.

\subsection*{Acknowledgements}

GT wishes to thank Z. Bajnok, L. Palla and also F. Smirnov for useful
discussions. This research was partially supported by the Hungarian
research funds OTKA T043582, K60040 and TS044839. GT was also supported
by a Bolyai J\'anos research scholarship.

\end{document}